\documentclass[10pt,preprint2,longnamesfirst]{aastex}
\received{2004 April~20}
\revised{2004 May~11}
\accepted{2004 May~12}
\journalid{xxx}{2004 August}
\paperid{204159}
\cpright{PD}{2004}
 
\shorttitle{Millisecond Pulsar Dust Disks} 
\shortauthors{Lazio et al.}

\begin{document} 
\title{Mid- and Far-Infrared ISO Limits on Dust Disks around Millisecond
	Pulsars}
\author{T.~Joseph~W.~Lazio \& J.~Fischer} 
\affil{Remote Sensing Division, Naval Research Laboratory,
         Washington, DC 20375-5351} 
\email{Joseph.Lazio@nrl.navy.mil} 
\email{Jacqueline.Fischer@nrl.navy.mil}

\begin{abstract}
We report 60 and~90~\micron\ observations of~7 millisecond pulsars
with \hbox{ISO}.  The pulsar \objectname[PSR]{PSR~B1257$+$12} is
orbited by three planets, and other millisecond pulsars may be orbited
by dust disks that represent planets that failed to form or their
residue.  We do not detect any infrared emission from the 7 pulsars in
our sample, and typical upper limits are 100~mJy.  Using a simple
model, we constrain the typical dust disk mass to be less than of order
100~M${}_\oplus$, assuming that the heating of any putative dust disk
would be coupled only weakly to the pulsar's emission.  If the planets around
\objectname[PSR]{PSR~B1257$+$12} are composed largely of metals, our
limits are probably an order of magnitude above plausible values for
the disk mass in metals.  Future observations with the Spitzer Space
Telescope should be capable of probing into the range of disk masses
that could plausibly give rise to planets.
\end{abstract}


\section{Introduction}\label{sec:intro}

The first extrasolar planets discovered were found around the
millisecond pulsar \objectname[PSR]{PSR~B1257$+$12} \citep{wf92}.  The
system consists of (at least) three planets, planet~A with
approximately a lunar mass, planet~B with $M = 4.3 \pm
0.2\,\mathrm{M}{}_\oplus$, and planet~C with $M = 3.9 \pm
0.2\,\mathrm{M}{}_\oplus$ \citep{kw03}.  Although planetary systems
around main-sequence stars had been long anticipated and numerous such
systems have been found since, pulsar planetary systems were
unexpected.  It was assumed that any planets orbiting the pulsar
progenitor would have become gravitationally unbound in the supernova
that produced the pulsar.

Various mechanisms have been proposed for the formation of these
planets \citep{ph92,mh01,h02}, but all generally rely on an accretion
disk around the pulsar within which the planets form.  Millisecond
pulsars are a class of pulsars that, subsequent to their formation,
undergo an episode of mass accretion from a companion
\citep{wv98}.  This process is thought to occur via an
accretion disk, which transfers angular momentum to the pulsar as
well, thereby spinning it up.  Various mechanisms exist to shut down
the accretion (e.g., evolution of the companion), but, if the
accretion is not 100\% efficient, the millisecond pulsar will be left
with an orbiting disk of material.  Such a residual accretion disk is
a natural location for the formation of planets.  Even if planets
form, the formation process may leave a debris disk.  \cite{gkw03}
investigated the long-term stability of a debris disk in the
\objectname[PSR]{PSR~B1257$+$12} system, finding a stable zone outside~1~\hbox{AU}.

Since the discovery of planets around \objectname[PSR]{PSR B1257$+$12}, a planet has
also been found around the pulsar \objectname[PSR]{PSR~B1620$-$26} \citep{bfs93,tat93,r94,jr97,tacl99,fjrz00,srhst03,rifh03}, in
the globular cluster M4.  In contrast to the planets orbiting
\objectname[PSR]{PSR~B1257$+$12}, which are thought to have formed \textit{in situ}, the
planet orbiting \objectname[PSR]{PSR~B1620$-$26} is thought to have been acquired during
a dynamical exchange within the globular cluster.

Pulsar planetary systems offer valuable insights, even if their total
number is unlikely ever to approach the number of planetary systems
around main-sequence stars.  Taken together the two pulsar planetary
systems already indicate that planets can form and exist in a wide variety of
environments.  The presence of terrestrial mass planets around
\objectname[PSR]{PSR~B1257$+$12} suggests that terrestrial planets may
be widespread, a hypothesis to be tested by future space missions such as Kepler and
the Terrestrial Planet Finder (TPF).  Planets orbiting main-sequence
stars near the Sun are found almost exclusively around stars with
solar- or super-solar metallicities \citep{g97,sim01}, which has led
to the belief that only stars with high metallicities can host
planets.  In contrast, the planet around \objectname[PSR]{PSR~B1620$-$26}, if it was
acquired during a dynamical exchange, probably has existed for a
substantial fraction of the age of the globular cluster M4.  This is a
low-metallicity globular cluster, suggesting that planets can form in
low-metallicity environments.

Although the notion that planets can form in a residual accretion
disk is plausible, no such examples of residual accretion disks are
known.  The presence of a planet (or stellar companion) can be
inferred using traditional pulsar timing techniques from the periodic
advance and delay of the arrival time of the pulsar's pulse, due to
the pulsar's reflex motion.  A relatively uniform disk of material
would produce little reflex motion and therefore would remain
undetected by these traditional techniques.  Detecting dust disks
around millisecond pulsars not only would elucidate the late stages of
millisecond pulsar ``spin up'' and planet formation, it would be a new
probe of the local environments around millisecond pulsars.

\begin{figure}
\epsscale{0.7}
\rotatebox{-90}{\plotone{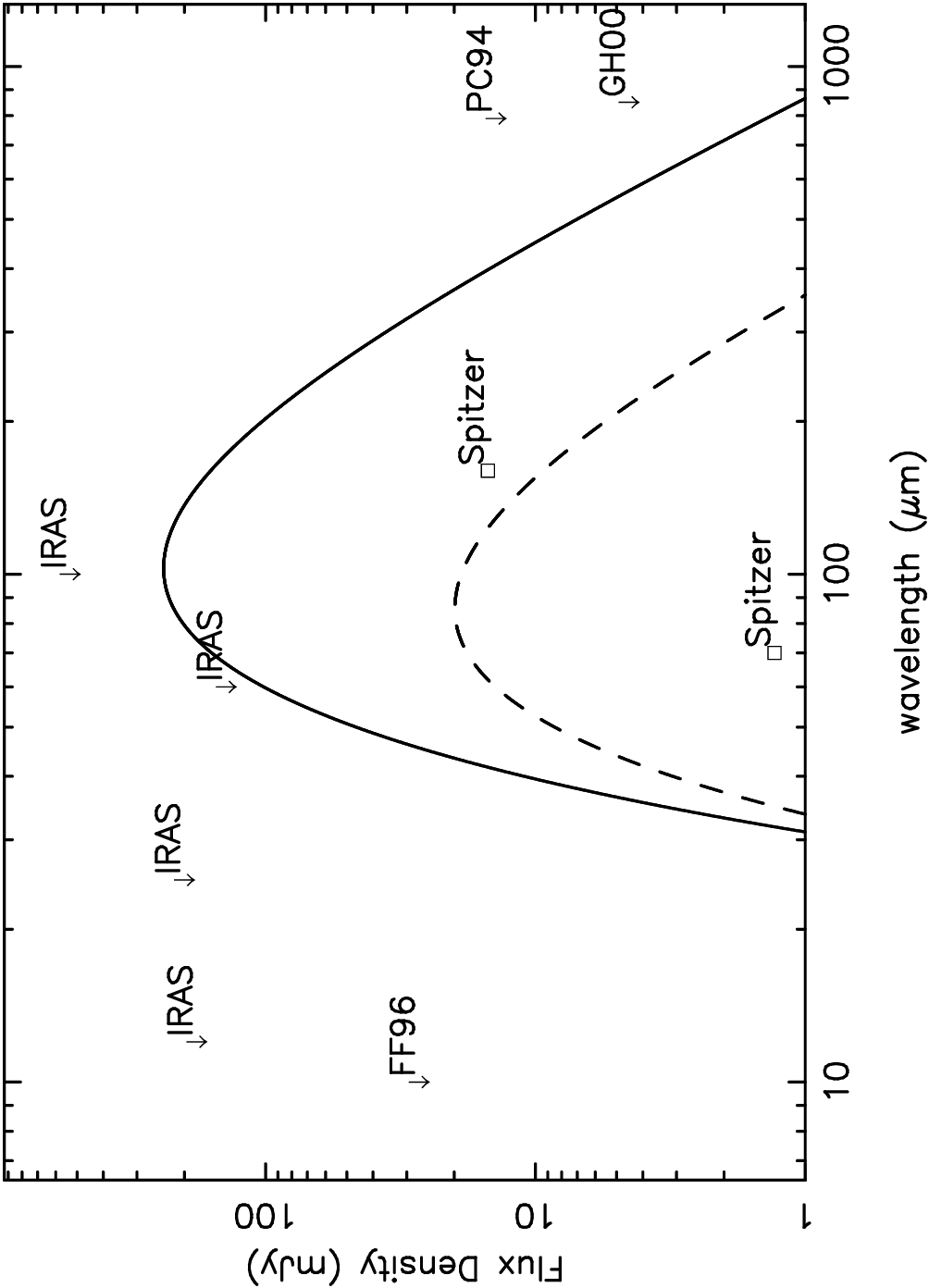}}
\caption[]{A summary of current limits on infrared dust emission from
around millisecond pulsars using
\protect\objectname[PSR]{PSR~B1257$+$12} as an example.  Shown are
limits placed by 10~\micron\ imaging by \cite[FF96]{ff96}, IRAS limits 
determined by \cite{ff96}, limits from~800~\micron\ imaging by
\cite[PC94]{pc94}, and limits from~850~\micron\ imaging by
\cite[GH00]{gh00}.  Also shown are the expected confusion limits for
the future observations with the Spitzer Space Telescope at~70
and~160~\micron.  This pulsar was not observed as part of this ISO
program.  The solid curve shows the expected emission from a
300~M${}_{\oplus}$ dust disk composed of dust particles 0.1~\micron\
in size and heated by 1\% of the spin-down luminosity of the pulsar
while the dotted curve shows the expected emission from a
10~M${}_{\oplus}$ disk composed of 0.1~\micron\ dust particles and
heated by 0.1\% of the spin-down luminosity \citep{ff96}.  Upper
limits on infrared emission for other millisecond pulsars are
similar.}
\label{fig:b1257+12}
\end{figure}

A modest number of unsuccessful searches for infrared emission from 
dust disks around millisecond pulsars have been conducted.
Figure~\ref{fig:b1257+12} summarizes the current situation using
\objectname[PSR]{PSR~B1257$+$12} as an example.  The limits for other
pulsars are similar.  

This paper reports 60 and~90~\micron\ observations of~7 pulsars with
the ISOPHOT instrument onboard the ISO satellite.  In
\S\ref{sec:observe} we describe the observations and present our
results and in \S\ref{sec:discuss} we describe how our results
constrain the presence of dust disks around millisecond pulsars and
present our conclusions.

\section{Observations and Data Analysis}\label{sec:observe}

We compiled a list of millisecond pulsars known prior to 1994~August
and with distances less than 1~kpc.  Distances are estimated from the
\cite{tc93} model and should be accurate to approximately 25\%.  Most
of these millisecond pulsars lie at high Galactic latitudes.  

Of these, seven were observed with the ISOPHOT instrument
\citep{lemkeetal96} onboard the ISO satellite \citep{kessleretal96}
between~1996 August and~1997 May.  Table~\ref{tab:log} summarizes the
observing details; we also report the distance to each pulsar and,
anticipating later discussion, whether or not it is a binary and its
spin-down luminosity.\footnote{%
The spin-down luminosity of a pulsar is a measure of its energy-loss
due to magnetic dipole radiation and is given by $L =
I\Omega\dot\Omega$, where $I$ is its moment of inertia and~$\Omega$ is 
its rotation frequency.
}
All of the observations used the P32 observing mode with the C100
detector.  In this mode the spacecraft was commanded to cover a series
of raster pointings around the nominal pulsar position.  At each
raster pointing an internal chopper pointed the beam toward~13
adjacent sky positions.  The throw of the chopper was larger than the
offset between raster pointings.  The result was that, in general, an
individual sky position within the raster was observed multiple times
or oversampled.  Before and after each observation of a pulsar, an
internal calibration source was observed.

\begin{deluxetable}{lcccccc}
\tablewidth{0pc}
\tablecaption{Pulsars Observed\label{tab:log}}
\tablehead{
                &                   &               & \colhead{Spin-down}
	&                      & 
	& \colhead{On-source} \\
 \colhead{Name} & \colhead{Binary?} & \colhead{Distance} & \colhead{Luminosity}
	& \colhead{Wavelength} & \colhead{P32 Raster}
	& \colhead{Time} \\
                &                   & \colhead{(kpc)}    & \colhead{(L${}_\odot$)} 
	& \colhead{(\micron)}  &                      & \colhead{(s)}
}

\startdata
\objectname[PSR]{PSR~J0034$-$0534} & Y & 0.98 & 10    & 90 & 3 $\times$ 8 & 1402 \\

\objectname[PSR]{PSR~J1640$+$2224} & Y & 1.19 &  0.88 & 60 & 3 $\times$ 6 & 1012 \\
                                   & &      &       & 90 & 3 $\times$ 8 & 848 \\

\objectname[PSR]{PSR~J1730$-$2304} & & 0.51 & $< 0.35$ & 60 & 3 $\times$ 6 & 1590 \\
                                   & &      &        & 90 & 3 $\times$ 8 & 1232 \\

\objectname[PSR]{PSR~B1855$+$09}   & Y & 0.91 &  1.1  & 60 & 3 $\times$ 6 & 1012 \\

\objectname[PSR]{PSR~J2124$-$3358} & & 0.25 &  1.7  & 60 & 3 $\times$ 6 & 1012 \\
                                   & &      &       & 90 & 3 $\times$ 8 & 848 \\

\\

\objectname[PSR]{PSR~J2145$-$0750} & Y & 0.50 & $< 0.048$ & 60 & 3 $\times$ 6 & 1590 \\
                                   &   &      &           & 90 & 3 $\times$ 8 & 848 \\
 
\objectname[PSR]{PSR~J2322$+$2057} & & 0.78 &  0.62 & 60 & 3 $\times$ 6 & 1804 \\
                                   & &      &       & 90 & 3 $\times$ 8 & 1402 \\

\enddata
\end{deluxetable}

The analysis of the pulsar observations largely followed the standard
ISOPHOT analysis pipeline.  The key difference was the amount of
``deglitching'' performed.  Glitches result from cosmic rays striking
the detector or secondary electrons produced by spacecraft materials
struck by primary cosmic rays.  Failure to remove glitches can corrupt
later calibration of \emph{all} data, not just of the portion
containing the glitches.
The standard ISOPHOT analysis pipeline removes glitches but does so
without making use of the redundancy implicit in the oversampled P32
observations.

Deglitching proceeded in the following fashion.  Within each
spacecraft pointing the chopper would sweep past a particular sky
position multiple times (typically 3--5 times).  For each sky
position, the median signal level was determined, then subtracted from
all observations at that sky position.  The observations from all sky
positions were then combined to form a signal strength histogram.  A
signal strength threshold was specified, and signals above this level
were eliminated.  Typically 3\%--10\% of the signals were eliminated
in this stage.  Depending upon the number of chopper sweeps per
spacecraft pointing and deglitching prior to this stage, the median
signal strength per sky position could not always be determined
accurately.  Thus, additional manual deglitching was done to remove
any remaining outlier signals.  Our use of the observations of the
internal calibration sources followed the standard ISOPHOT analysis
pipeline.

After deglitching and calibration using the internal calibration
sources, mapping was done within the ISOPHOT Interactive Analysis
package.  Measurements from the individual detector pixels were
co-added to form a sky image, with the contributions from the individual 
detector pixels weighted by their distances from the image pixels.
Doing so takes into account the beam profile falling on each detector
pixel.  We also employed a median flat field, which has the effect of
reducing substantially our sensitivity to any extended emission in the 
field.  As we are attempting to detect point sources, we regard 
this reduced sensitivity to extended emission as unimportant.

In no case have we identified a source at the location of a pulsar.
Utilizing the inner quarter of the image, we determined the rms noise
level.  We take our upper limits to be 3 times this rms noise level.
Table~\ref{tab:limits} summarizes the upper limits.

\begin{deluxetable}{lc}
\tablewidth{0pc}
\tablecolumns{2}
\tablecaption{Dust Disk IR Emission Upper Limits\label{tab:limits}}
\tablehead{
 \colhead{Name} & \colhead{Flux Density} \\
                & \colhead{(mJy)}}

\startdata
\cutinhead{60~\micron}

\objectname[PSR]{PSR~J1640$+$2224} & 80 \\
\objectname[PSR]{PSR~J1730$-$2304} & 35 \\ 
\objectname[PSR]{PSR~B1855$+$09}   & 190 \\
\objectname[PSR]{PSR~J2124$-$3358} & 100 \\
\objectname[PSR]{PSR~J2145$-$0750} & 100 \\

\objectname[PSR]{PSR~J2322$+$2057} & 59 \\

\cutinhead{90~\micron}

\objectname[PSR]{PSR~J0034$-$0534} & 48 \\ 
\objectname[PSR]{PSR~J1640$+$2224} & 59 \\ 
\objectname[PSR]{PSR~J1730$-$2304} & 140 \\
\objectname[PSR]{PSR~J2124$-$3358} & 55 \\
\objectname[PSR]{PSR~J2145$-$0750} & 73 \\

\objectname[PSR]{PSR~J2322$+$2057} & 39 \\

\enddata
\tablecomments{Upper limits are 3$\sigma$.}
\end{deluxetable}

\section{Discussion and Conclusions}\label{sec:discuss}

We have not detected infrared emission associated with any of the
pulsars observed with \hbox{ISO}.  Other infrared
and sub-millimeter observations of millisecond pulsars have been
conducted, and all of these have yielded only upper limits as
well.  Those observations most relevant to our sample of millisecond
pulsars are those by \cite{ff96} at~10~\micron\ and \cite{gh00}
at~850~\micron.  \cite{ff96} also utilized IRAS observations to obtain
upper limits on the infrared emission from their sample of pulsars.
As Figure~\ref{fig:b1257+12} shows, the upper limits set by
IRAS are typically well above the limits set by our ISO observations.
Moreover, there is unfortunately little overlap between these three
samples of pulsars (those whose observations are reported here,
\citealt*{ff96}, and \citealt*{gh00}).  Most of the pulsars that have
been observed between~10 and~850~\micron\ have been observed at only
one or two wavelengths.

\cite{ff96} developed a model for the infrared emission from a dust
disk around a millisecond pulsar.  Their model assumes that the disk
consists of particles of a uniform radius~$a$ heated by a
fraction~$f_{\mathrm{sd}}$ of the pulsar's spin-down
luminosity~L${}_{\mathrm{sd}}$.  The total mass of the disk is $m_d$.
While the model is simplistic---an actual dust disk presumably
consists of particles with a range of sizes, the heating mechanism is
left unspecified, non-equilibrium effects such as stochastic heating
are ignored, and the impact of any stellar companions (see
Table~\ref{tab:log}) on the disk are ignored---we believe that this
simplicity is justified given the uncertainties of the heating
mechanism and of the environs of a millisecond pulsar.

In this model, for $f_{\mathrm{sd}} \sim 1$\%, typical dust
temperatures are predicted to be $T \approx 10$--50~K for disks having
$m_d \sim 100$~M${}_\oplus$ and~$a \sim 1$~\micron\ and heated by a
pulsar with L${}_{\mathrm{sd}} \sim 1$~L${}_\odot$.  These
temperatures are similar to the lower temperature range used by
\cite{k-mhppns02} and considerably lower than those assumed ($\approx
150$~K) by \cite{pc94}, who estimated disk temperatures by scaling
from observations of T~Tauri stars.  The lower temperatures result
from our assumption of a weaker coupling between the pulsar's
spin-down luminosity and the disk.  \cite{pc94} considered disk
temperature to be a major uncertainty in converting from measured flux
densities to inferred disk masses.  Accordingly, our assumption of a
weaker coupling means that larger disk masses can be tolerated without
violating the observational constraints.

\begin{figure}[tbh]
\epsscale{0.7}
\rotatebox{-90}{\plotone{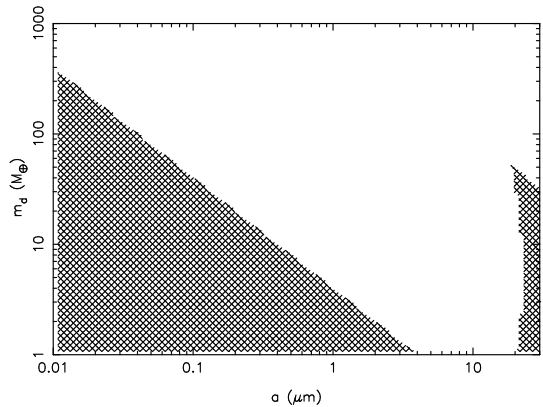}}
\caption{Allowed disk masses and grain sizes for a dust disk orbiting
\protect\objectname{PSR~J0034$-$0534}.  The cross-hatched region
indicates disk masses and grain sizes that do not violate the
observational limits at~90 and~850~\micron.  A heating efficiency of
$f_{\mathrm{sd}} = 1$\% has been assumed.}
\label{fig:like2}
\end{figure}

Given the paucity of data, it is not possible, in general, to
constrain all three parameters of this model with the existing
observations.  We therefore adopt an approach in which we infer limits
on two parameters of the Foster \& Fischer~(1996) model for fiducial
values of the third parameter.  Here, as an example, we consider the
millisecond pulsar \objectname{PSR~J0034$-$0534} \citep{bailesetal94}
which has a probable white dwarf companion, is at a distance of~1~kpc,
and has a spin-down luminosity of~10~L${}_\odot$.  Greaves \&
Holland~(2000) placed a $2\sigma$ limit of~3.7~mJy at 850~\micron, and
we place a $2\sigma$ limit of~50~mJy at~90~\micron.
Figure~\ref{fig:like2} shows the allowed region of the disk mass-grain
size plane given these observational limits and an assumed heating
efficiency of $f_{\mathrm{sd}} = 1$\%.

Allowed regions in the $m_d$-$a$ plane occur for one of two possible
reasons.  First, the peak of the dust disk emission may appear
shortward of~90~\micron, where no constraints exist for this pulsar,
with the Rayleigh-Jeans tail of the emission falling below the two
measured values.  This region is to the lower left in
Figure~\ref{fig:like2}.  Second, the peak of the emission may
appear between~90~\micron\ and~850~\micron, but with a magnitude
comparable to that measured at~90~\micron\ so that the Rayleigh-Jeans
tail again does not violate the 850~\micron\ limit while the Wien tail
of the emission does not violate the 90~\micron\ limit.  This region
is to the lower right in Figure~\ref{fig:like2}.  Obviously, a
lower value of~$f_{\mathrm{sd}}$ would produce larger allowed regions
in the $m_d$-$a$ plane.  Larger allowed regions would also exist for
other pulsars (Table~\ref{tab:log}) with smaller spin-down luminosities.

We conclude that, with the current observational constraints and the
assumption of a fairly weak energy coupling between pulsars and disks, 
dust disks of order 100~M${}_\oplus$ easily could exist around
millisecond pulsars.  \cite{k-mhppns02} have reached similar
conclusions based on~15 and~90~\micron\ observations.  To the extent that
such disks would be uniform, they would also escape detection from
traditional pulsar timing techniques.  Pulsar timing techniques
utilize the advance or delay of the pulse arrival time resulting from
the pulsar's reflex motion to detect planetary or stellar companions.
A relatively uniform dust disk would produce little reflex
motion.\footnote{%
One possible exception to this conclusion would be a dust disk
illuminated directly by the pulsar's beam.  In this case the
relativistic particle flow from the pulsar beam potentially could
ionize portions of the disk and induce plasma propagation delays which
might be detectable.
}

The current limits on dust disk masses are far larger than the mass of
the disk thought to have produced the planets around
\objectname[PSR]{PSR~B1257$+$12}.  The minimum combined mass of the
two larger planets in that system is 8.2~M${}_\oplus$ \citep{kw03}.
Assuming that these planets are composed largely of metals, we expect
that any dust mass prior to the planets' formation would be comparable
in magnitude.  Indeed, \cite{h02} has shown how an initial disk of
mass 0.1--$10^{-3}$~M${}_\odot$ containing of order 10~M${}_\oplus$ in
metals could form a system similar to that orbiting
\objectname[PSR]{PSR~B1257$+$12}.  (See also \citealt*{mh01} and
Figure~\ref{fig:b1257+12}.)  For those pulsars orbited by stellar
companions, the companions will introduce regions of limited orbital
stability within the disks, potentially implying even smaller expected
disk masses.  We conclude that current observational limits on dust
disk masses are at least an order of magnitude above plausible values.

The mid- to far-infrared detectors (24, 70, and~160~\micron) on the
Spitzer Space Telescope should have sensitivities some 1--2 orders of
magnitude better than the limits we report here.  At a minimum, we
expect that future observations with the Spitzer Space Telescope may
require more sophisticated modelling of disks, including the possible
effects of stellar companions.  If future observations with the
Spitzer Space Telescope do not detect infrared emission from dust
disks around millisecond pulsars, the resulting mass limits should be
in the range of~10~M${}_\oplus$, sufficient to begin placing stringent
constraints on their existence, or the temperatures of any pulsar dust
disks must be no more than a few Kelvin.

\acknowledgements

We thank the organizers of the ISOPHOT Workshop on PHT32 Oversampled
Mapping, particularly R.~Tuffs, C.~Gabriel, N.~Lu, and B.~Schulz for
their many helpful discussions, and R.~Tuffs for his deglitching
software.  Without their assistance, no results would be reported
here.  We thank the referee for comments that helped us clarify
certain points and C.~Chandler for helpful discussions.  The results
reported here are based on observations with ISO, an ESA project with
instruments funded by ESA Member States (especially the PI countries:
France, Germany, the Netherlands and the United Kingdom) and with the
participation of ISAS and NASA.  The ISOPHOT data presented in this
paper were reduced using PIA, which is a joint development by the ESA
Astrophysics Division and the ISOPHOT consortium, with the
collaboration of the Infrared Analysis and Processing Center (IPAC)
and the Instituto de Astrof{\'\i}sica de Canarias (IAC).  Basic
research in astronomy at the NRL is supported by the Office of Naval
Research.


\begin{thebibliography}{}
\bibitem[\protect\citeauthoryear{Backer, Foster, \& Sallmen}{Backer et 
	al.}{1993}]{bfs93}
	Backer, D.~C., Foster, R.~S., \& Sallmen, S.  1993, Nature, 365, 817

\bibitem[\protect\citeauthoryear{Bailes et al.}{1994}]{bailesetal94}
	Bailes, M., et al.  1994, ApJ, 425, L41

\bibitem[\protect\citeauthoryear{Ford et al.}{2000}]{fjrz00}
	Ford, E.~B., Joshi, K.~J., Rasio, F.~A., \& Zbarsky, B.
	2000, \apj, 528, 336

\bibitem[\protect\citeauthoryear{Foster \& Fischer}{1996}]{ff96}
	Foster, R.~S.\ \& Fischer, J.  1996, \apj, 460, 902

\bibitem[\protect\citeauthoryear{Gozdziewski et al.}{2003}]{gkw03}
        Gozdziewski, K., Konacki, M., \& Wolszczan, A.  2003, ApJ,
        submitted; astro-ph/0310750

\bibitem[\protect\citeauthoryear{Gonzalez}{1997}]{g97}
        Gonzalez, G.  1997, MNRAS, 285, 403

\bibitem[\protect\citeauthoryear{Greaves \& Holland}{2000}]{gh00}
	Greaves, J.~S.\ \& Holland, W.~S.  2000, \mnras, 316, L21

\bibitem[\protect\citeauthoryear{Hansen}{2002}]{h02}
	Hansen, B.~M.~S.  2002, in Stellar Collisions, Mergers, and
	Their Consequences, ed.~M.~M.Shara (San Francisco: ASP) p.~221

\bibitem[\protect\citeauthoryear{Joshi \& Rasio}{1997}]{jr97}
	Joshi, K.~J.\ \& Rasio, F.~A.  1997, \apj, 479, 948

\bibitem[\protect\citeauthoryear{Kessler et al.}{1996}]{kessleretal96} 
	Kessler, M.~F., et al.  1996, \aap, 315, L27  

\bibitem[\protect\citeauthoryear{Koch-Miramond et
	al.}{2002}]{k-mhppns02}
	Koch-Miramond, L., Haas, M., Pantin, E., Podsiadlowski, Ph.,
	Naylor, T., \& Sauvage, M.  2002, \aap, 387, 233

\bibitem[\protect\citeauthoryear{Konacki \& Wolszczan}{2003}]{kw03}
	Konacki, M.\ \& Wolszczan, A.  2003, ApJ, 591, L147

\bibitem[\protect\citeauthoryear{Lemke et al.}{1996}]{lemkeetal96}
	Lemke, D., et al.  1996, \aap, 315, L64 

\bibitem[\protect\citeauthoryear{Miller \& Hamilton}{2001}]{mh01}
	Miller, M.~C.\ \& Hamilton, D.~P.  2001, ApJ, 550, 863

\bibitem[\protect\citeauthoryear{Phillips \& Chandler}{1994}]{pc94} 
	Phillips, J.~A.\ \& Chandler, C.~J.  1994, \apj, 420, L83

\bibitem[\protect\citeauthoryear{Phinney \& Hansen}{1992}]{ph92}
	Phinney, E.~S.\ \& Hansen, B.~M.~S.  1992, in Planets around
	Pulsars, eds.\  J.~A.~Phillips, S.~E.~Thorsett, \&
	S.~R.~Kulkarni (San Francisco: ASP) p.~371

\bibitem[\protect\citeauthoryear{Rasio}{1994}]{r94}
	Rasio, F.~A.  1994, \apj, 427, L107

\bibitem[\protect\citeauthoryear{Richer et al.}{2003}]{rifh03}
        Richer, H.~B., Ibata, R., Fahlman, G.~G., \& Huber, M.  2003,
        ApJ, 597, L45

\bibitem[\protect\citeauthoryear{Santos et al.}{2001}]{sim01}
        Santos, N.~C., Israelian, G., \& Mayor, M.  2001, A\&A, 373, 1019

\bibitem[\protect\citeauthoryear{Sigurdsson et al.}{2003}]{srhst03}
        Sigurdsson, S., Richer, H.~B., Hansen, B.~M., Stairs, I.~H.,
        \& Thorsett, S.~E.  2003, Science, 301, 193

\bibitem[\protect\citeauthoryear{Taylor \& Cordes}{1993}]{tc93}
	Taylor, J.~H.\ \& Cordes, J.~M.  1993, \apj, 411, 674

\bibitem[\protect\citeauthoryear{Thorsett et al.}{1999}]{tacl99}
	Thorsett, S.~E., Arzoumanian, Z., Camilo, F., \& Lyne, A.~G.
	1999, \apj, 523, 763

\bibitem[\protect\citeauthoryear{Thorsett, Arzoumanian, \&
	Taylor}{Thorsett et al.}{1993}]{tat93}
	Thorsett, S.~E., Arzoumanian, Z., \& Taylor, J.~H. 
	1993, \apj, 412, L33

\bibitem[\protect\citeauthoryear{van~den~Heuvel}{1995}]{v95}
	van~den~Heuvel, E.~P.~J.  1995, J.\ Astrophys.\ Astron., 16, 255

\bibitem[\protect\citeauthoryear{Wijnands \&
        van~der~Klis}{1998}]{wv98}
        Wijnands, R.\ \& van~der~Klis, M.  1998, Nature, 394, 344

\bibitem[\protect\citeauthoryear{Wolszczan \& Frail}{1992}]{wf92}
	Wolszczan, A.\ \& Frail, D.~A.  1992, \nat, 355, 145
\end{thebibliography}
\end{document}